\documentclass[amsmath,aps,showpacs,a4paper,10pt]{revtex4}
 \usepackage{epsf}
 \usepackage{graphicx}    % Include figure files

\begin{document}

 \newcommand{\be}[1]{\begin{equation}\label{#1}}
 \newcommand{\ee}{\end{equation}}
 \newcommand{\bea}{\begin{eqnarray}}
 \newcommand{\eea}{\end{eqnarray}}
 \def\disp{\displaystyle}

 \def\gsim{ \lower .75ex \hbox{$\sim$} \llap{\raise .27ex \hbox{$>$}} }
 \def\lsim{ \lower .75ex \hbox{$\sim$} \llap{\raise .27ex \hbox{$<$}} }

 \begin{titlepage}

 \begin{flushright}
 arXiv:1605.04571
 \end{flushright}

 \title{\Large \bf Cosmological Constant, Fine Structure
 Constant and Beyond}

 \author{Hao~Wei\,}
 \email[\,email address:\ ]{haowei@bit.edu.cn}
 \affiliation{School of Physics,
 Beijing Institute of Technology, Beijing 100081, China}

 \author{Xiao-Bo~Zou\,}
 \affiliation{School of Physics,
 Beijing Institute of Technology, Beijing 100081, China}

 \author{Hong-Yu~Li\,}
 \affiliation{School of Physics,
 Beijing Institute of Technology, Beijing 100081, China}

 \author{Dong-Ze~Xue\,}
 \affiliation{School of Physics,
 Beijing Institute of Technology, Beijing 100081, China}

 \begin{abstract}\vspace{1cm}
 \centerline{\bf ABSTRACT}\vspace{2mm}
 In the present work, we consider the cosmological constant
 model $\Lambda\propto\alpha^{-6}$, which is well motivated
 from three independent approaches. As is well known, the hint
 of varying fine structure constant $\alpha$ was found in 1998.
 If $\Lambda\propto\alpha^{-6}$ is right, it means that the
 cosmological constant $\Lambda$ should also be varying. Here,
 we try to develop a suitable framework to model this varying
 cosmological constant $\Lambda\propto\alpha^{-6}$, in which
 we view it from an interacting vacuum energy perspective.
 Then, we consider the observational constraints on these
 models by using the 293 $\Delta\alpha/\alpha$ data from the
 absorption systems in the spectra of distant quasars. We find
 that the model parameters can be tightly constrained to the
 very narrow ranges of ${\cal O}(10^{-5})$ typically. On the
 other hand, we can also view the varying cosmological constant
 model $\Lambda\propto\alpha^{-6}$ from another perspective,
 namely it can be equivalent to a model containing ``dark
 energy'' and ``warm dark matter'', but there is no interaction
 between them. We find that this is also fully consistent
 with the observational constraints on warm dark matter.
 \end{abstract}


 \maketitle

 \end{titlepage}

 \renewcommand{\baselinestretch}{1.0}

%============================= section 1 ===================================

\section{Introduction}\label{sec1}

The cosmological constant has been one of the long-standing
 issues in physics and cosmology since it was introduced by
 Einstein in 1917~\cite{r1} for a static universe. However,
 Hubble discovered in 1929~\cite{r2} that the universe is
 expanding. Then, Einstein abandoned the cosmological constant
 as the ``biggest blunder'' of his life~\cite{r3}. From 1929
 until the early 1990s, most physicists and cosmologists
 assumed the cosmological constant to be zero. Since the vacuum
 energy is equivalent to the cosmological constant~\cite{r4},
 an exactly zero cosmological constant requires that the bare
 cosmological constant should be exactly canceled by the vacuum
 energy. This is a difficult problem~\cite{r5} (sometimes it is
 called the (old) cosmological constant problem in the literature).
 In 1998, the accelerated expansion of the universe was
 discovered~\cite{r6}, and since then dark energy has been one
 of the most active fields in cosmology~\cite{r7,r8}. So, the
 cosmological constant was revived again, since the simplest
 candidate of dark energy is a tiny positive cosmological constant.
 However, it is difficult to understand why the observable
 cosmological constant is about 120 orders of magnitude smaller than
 its natural expectation of the vacuum energy~\cite{r5,r7}. Now, the
 (new) cosmological constant problem becomes the question why
 the non-zero cosmological constant is so tiny. It means that a
 fine-tuning is necessary when the bare cosmological constant
 is canceled by the vacuum energy~\cite{r5,r7}. In fact, the
 cosmological constant is still an important topic in physics
 and~cosmology~by~now.

It is commonly believed that the cosmological constant problem
 can only be solved ultimately in a unified theory of quantum
 gravity and the standard model of electroweak and strong
 interactions, which is still absent so far. Nevertheless,
 many attempts have been made in the literature. One of the
 interesting ideas is the so-called axiomatic approach to the
 cosmological constant~\cite{r9}. In this approach, the
 cosmological constant is derived from four axioms, but the
 underlying physical origin (say, the theory of quantum
 gravity) is still unknown. It is proposed in close analogy to
 the Khinchin axioms in information theory. The well-known
 Khinchin axioms can uniquely derive the Shannon entropy, on
 which the entire mechanism of statistical mechanics is based
 (see the textbook e.g.~\cite{r10}).

The Khinchin axioms in information theory~\cite{r11} describe
 the most desirable properties an information measure $I$
 should have. Axiom {\bf K1} ``fundamentality'': an information
 measure $I$ only depends on the probabilities $p_i$ (the
 fundamental quantities) of the events under consideration and
 nothing else. Axiom {\bf K2} ``boundedness'': there is a
 lower bound for the information measure $I$. Axiom {\bf K3}
 ``simplicity'': the information measure $I$ should take the
 simplest description. Axiom {\bf K4} ``invariance'': there is
 a suitable scale transformation in the space of probabilities
 and information measures that leaves the physical contents
 invariant. These four Khinchin axioms look very natural and
 simple. However, from such natural and simple axioms, one can
 uniquely fix the functional form of the Shannon information
 which is extremely important for the statistical
 mechanics~\cite{r10}.

Inspired by the successful Khinchin axiomatic approach to the
 Shannon entropy in information theory, Beck~\cite{r9} proposed
 four axioms in close analogy to the Khinchin axioms. Axiom
 {\bf B1} ``fundamentality'': the cosmological constant
 $\Lambda$ only depends on fundamental constants of nature.
 Axiom {\bf B2} ``boundedness'': the cosmological constant is
 bounded from below, $\Lambda>0$. Axiom {\bf B3}
 ``simplicity'': the cosmological constant $\Lambda$ is given by the
 simplest possible formula consistent with the other axioms.
 Axiom {\bf B4} ``invariance'': the cosmological constant $\Lambda$
 formed with potentially different values of fundamental
 parameters leaves the large-scale physics of the universe
 scale invariant. These four axioms look also very natural and
 simple. From these natural and simple axioms, Beck~\cite{r9}
 derived the explicit form of the cosmological constant,
 \be{eq1}
 \Lambda=
 \frac{G^2}{\hbar^4}\left(\frac{m_e}{\alpha}\right)^6\,,
 \ee
 where $\alpha$ is the fine structure constant, $G$ is the
 gravitational constant, $\hbar$ is the reduced Planck
 constant, and $m_e$ is the electron mass. Accordingly, the
 (observable) vacuum energy density is given by~\cite{r9}
 \be{eq2}
 \rho_\Lambda=\frac{c^4}{8\pi G}\,\Lambda=\frac{G}{8\pi}
 \frac{c^4}{\hbar^4}\left(\frac{m_e}{\alpha}\right)^6\,,
 \ee
 where $c$ is the speed of light. Numerically, this formula
 yields $\rho_\Lambda\simeq 4.0961\,{\rm GeV/m^3}$, which
 can pass the current observational constraints with flying
 colors. We refer to~\cite{r9} for the detailed derivations.

In fact, Beck~\cite{r9} is not the first and the only one who
 derived the cosmological constant given in Eq.~(\ref{eq1}).
 It was independently derived from other arguments in the
 literature. Using the generalized Buchdahl identity, Boehmer
 and Harko~\cite{r12} argued that the existence
 of a non-negative $\Lambda$ imposes a lower bound on the
 mass $M$ and density $\rho$ for general relativistic objects
 with radius $\cal R$,
 \be{eq3}
 2GM\geq\frac{\Lambda c^2}{6}{\cal R}^3\,,~~~~~~~
 \rho=\frac{3M}{4\pi{\cal R}^3}\geq
 \frac{\Lambda c^2}{16\pi G}\,.
 \ee
 On the other hand, Wesson~\cite{r13} argued that the mass
 is quantized according to the rule
 $m=(n\hbar/c)\sqrt{\Lambda/3}$, and the minimum
 mass corresponding to the ground state $n=1$ is given by
 \be{eq4}
 m_P=\frac{\hbar}{c}\sqrt{\frac{\Lambda}{3}}\,,
 \ee
 which is indeed a very small mass. Boehmer and
 Harko~\cite{r12,r14} proposed to identify the minimum mass
 in Eq.~(\ref{eq3}) with the one in Eq.~(\ref{eq4}), and
 found that the radius corresponding to $m_P$ is given by
 \be{eq5}
 {\cal R}_P=48^{1/6} \left(\frac{\hbar G}{c^3}\right)^{1/3}
 \Lambda^{-1/6} \simeq 1.9\,\ell_{pl}^{2/3}\Lambda^{-1/6}\,,
 \ee
 where $\ell_{pl}$ is the Planck length. Noting the radius
 ${\cal R}_P$ is of the same order of magnitude as the
 classical radius of the electron $r_e=e^2/(m_e c^2)$ (where
 $e$ is the electron charge), Boehmer and Harko~\cite{r14}
 further proposed to formally equate ${\cal R}_P$ with $r_e$
 while the term of ${\cal O}(1)$ is neglected, and then
 they found that the cosmological constant is
 given by~\cite{r14}
 \be{eq6}
 \Lambda=\frac{\ell_{pl}^4}{r_e^6}=
 \frac{\hbar^2 G^2 m_e^6 c^6}{e^{12}}=\frac{G^2}{\hbar^4}
 \left(\frac{m_e}{\alpha}\right)^6\,,
 \ee
 in which we have used the definition $\alpha=e^2/(\hbar c)$.
 Clearly, the same result given in Eq.~(\ref{eq1}) has been
 independently derived from completely different arguments.

The third independent approach to derive the cosmological
 constant $\Lambda$ in Eq.~(\ref{eq1}) is the well-known
 Eddington-Dirac large number hypothesis~\cite{r15,r16,r17}.
 Nottale in 1993~\cite{r18} (see also~\cite{r9}) has written down a
 large number hypothesis connecting cosmological parameters
 with standard model parameters,
 \be{eq7}
 \alpha\frac{m_{pl}}{m_e}=\left(
 \frac{\Lambda^{-1/2}}{\ell_{pl}}\right)^{1/3}\,,
 \ee
 where $m_{pl}$ is the Planck mass. It is easy to check
 that Eq.~(\ref{eq7}) is equivalent to Eqs.~(\ref{eq1}) and
 (\ref{eq6}) in fact.

We note that the cosmological constant $\Lambda$ given in
 Eq.~(\ref{eq1}) is related to the fine structure constant
 $\alpha$ according to $\Lambda\propto\alpha^{-6}$. This is
 interesting for us. As is well known, in the same year 1998
 when the accelerated cosmic expansion was discovered, the
 evidence for cosmological evolution of the fine structure
 constant $\alpha$ has also been found~\cite{r19,r20}. Using
 the absorption systems in the spectra of distant quasars,
 Webb~{\it et al.}~\cite{r19} found the first evidence for the
 time variation of $\alpha$, namely $\Delta\alpha/\alpha\equiv
 (\alpha-\alpha_0)/\alpha_0=(-1.1\pm 0.4)\times 10^{-5}$ over
 the redshift range $0.5<z<1.6$, where $\alpha_0$ is the
 present value of $\alpha$. Three years later, they improved
 the evidence to $4\sigma$, namely $\Delta\alpha/\alpha=(-0.72
 \pm 0.18)\times 10^{-5}$ over the redshift range
 $0.5<z<3.5$~\cite{r20}. The fine structure constant $\alpha$
 was smaller in the past, and it is not a true constant in
 fact. Nowadays, a time-varying $\alpha$ has been extensively
 discussed in the community. There are many works on this topic
 in the literature~\cite{r21,r22,r23,r60}. If the cosmological
 constant $\Lambda$ given in Eq.~(\ref{eq1}) is right, it
 should also be time-varying,
 because $\Lambda\propto\alpha^{-6}$. In the literature
 (e.g.~\cite{r24,r25}), there exist some $\Lambda(t)$ models
 already. However, most of them are written by
 hand, e.g. $\Lambda\propto H^2$, $\Lambda\propto \ddot{a}/a$,
 $\Lambda\propto R_{sc}\,$, $\Lambda\propto\rho_m$, where $H$
 is the Hubble parameter, $a$ is the scale factor, $R_{sc}$
 is the scalar curvature, $\rho_m$ is the density of matter.
 Different from the $\Lambda(t)$ models purely written by
 hand, the time-varying $\Lambda\propto\alpha^{-6}$ given in
 Eq.~(\ref{eq1}) is well motivated, as is shown above.

In the present work, we are interested to study the varying
 cosmological constant $\Lambda\propto\alpha^{-6}$. In
 Sec.~\ref{sec2}, we try to develop a suitable framework to
 model the varying cosmological constant. In Sec.~\ref{sec3},
 we consider the observational constraints on the varying
 $\Lambda$ models. In Sec.~\ref{sec4}, the possible connection
 between the varying cosmological constant and warm dark matter
 is discussed. In Sec.~\ref{sec5}, some brief concluding
 remarks are given.

%============================= section 2 ===================================

\section{Varying cosmological constant and fine structure constant}\label{sec2}

Here, we try to develop a suitable framework to model the
 varying cosmological constant $\Lambda$ given in
 Eq.~(\ref{eq1}). For convenience, we instead use the vacuum
 energy density $\rho_\Lambda$ given in Eq.~(\ref{eq2}), which
 is equivalent to $\Lambda$ in fact. Throughout this work, we
 use the terms ``cosmological constant'' and ``vacuum energy''
 interchangeably. If the cosmological constant is
 varying, we have $\dot{\rho}_\Lambda=-Q\not=0$, where a dot
 denotes the derivative with respect to cosmic time $t$. To
 preserve the total energy conservation equation
 $\dot{\rho}_{tot}+3H(\rho_{tot}+p_{tot})=0$, a coupling
 between the vacuum energy and the pressureless matter is
 necessary, and hence $\dot{\rho}_m+3H\rho_m=Q\not=0$, where
 $\rho_m$ is the density of pressureless matter,
 $\rho_{tot}=\rho_\Lambda+\rho_m$, and $p_{tot}$ is the total
 pressure. Note that the equation-of-state parameter (EoS) of
 the cosmological constant $w_\Lambda=-1$, and the EoS of the
 pressureless matter $w_m=0$. Throughout this work, we
 assume that only the fine structure ``constant'' $\alpha$ is
 varying, and all the other fundamental constants $\hbar$,
 $G$, $c$, $m_e$ are true constants, i.e. they do not vary
 indeed. Since $\alpha=e^2/(\hbar c)$, this means that only
 the electron charge $e$ is varying. Therefore, we have
 $\rho_\Lambda\propto\Lambda\propto\alpha^{-6}$. It is easy to
 see that
 \be{eq8}
 \frac{\dot{\rho}_\Lambda}{\rho_\Lambda}=
 -6\frac{\dot{\alpha}}{\alpha}\,,
 \ee
 and then the total energy conservation equation
 can be preserved according to
 \bea
 &&\dot{\rho}_\Lambda=-Q=-6\frac{\dot{\alpha}}{\alpha}
 \rho_\Lambda\,,\label{eq9}\\[1mm]
 &&\dot{\rho}_m+3H\rho_m=Q=6\frac{\dot{\alpha}}{\alpha}
 \rho_\Lambda\,.\label{eq10}
 \eea
 The coupling term $Q=6\rho_\Lambda\dot{\alpha}/\alpha\not=0$
 if the fine structure ``constant'' $\alpha$ is varying. In
 this work, we consider a spatially flat Friedmann-Robertson-Walker
 (FRW) universe containing only the vacuum energy and the
 pressureless matter. $H\equiv\dot{a}/a$ is the Hubble
 parameter, and $a=(1+z)^{-1}$ is the scale factor (we have set
 $a_0=1$; the subscript ``0'' indicates the present value of
 corresponding quantity; $z$ is the redshift). In this way, we
 can turn the varying cosmological constant model into an
 interacting vacuum energy model. The vacuum energy interacts
 with the pressureless matter by exchanging energy between
 them.

Due to the interaction $Q$, the evolutions of $\rho_m$
 and $\rho_\Lambda$ should deviate from the ones without
 interaction, namely $\rho_m\propto a^{-3}$ and
 $\rho_\Lambda=const.$ If the coupling term $Q$ is given, one
 can derive the evolutions of $\rho_m$ and $\rho_\Lambda$.
 However, the logic can be reversed. If the deviated evolutions
 of $\rho_m$ and/or $\rho_\Lambda$ are given, we can find
 the corresponding interaction $Q$ from Eqs.~(\ref{eq9}),
 (\ref{eq10}), and then the evolution of $\alpha$. Inspired by
 e.g.~\cite{r24,r26,r27,r28,r29}, we consider two different
 types of models to characterize the deviated evolutions of
 $\rho_m$ and/or $\rho_\Lambda$ in the following
 two subsections, respectively.

%============================= section 2.1 ===================================

\subsection{Type I models}\label{sec2a}

Inspired by e.g.~\cite{r27,r28,r29}, the type I models are
 characterized by
 \be{eq11}
 \frac{\rho_\Lambda}{\rho_m}=f(a)\,,
 \ee
 where $f(a)$ can be any function of scale factor $a$. If
 $f(a)\propto a^3$, it corresponds to $\Lambda$CDM model whose
 $\rho_\Lambda=const.$ and $\rho_m\propto a^{-3}$. From
 Eq.~(\ref{eq11}) and Friedmann
 equation $H^2=8\pi G(\rho_\Lambda+\rho_m)/3$, we have
 \be{eq12}
 \Omega_\Lambda = \frac{f}{1+f}\,,~~~~~~~~
 \Omega_m = \frac{1}{1+f}\,,
 \ee
 where $\Omega_i\equiv 8\pi G\rho_i/(3H^2)$ ($i=\Lambda,\,m$)
 are the fractional energy densities of the vacuum energy and
 matter, respectively. Substituting $\rho_\Lambda=\rho_m f(a)$
 into Eq.~(\ref{eq9}) and using $\dot{\rho}_m$
 from Eq.~(\ref{eq10}), we find
 \be{eq13}
 Q=-H\rho_m\Omega_\Lambda\left(a\frac{f^\prime}{f}-3\right)
 =-H\rho_\Lambda\Omega_m\left(a\frac{f^\prime}{f}-3\right)\,,
 \ee
 where a prime denotes the derivative with respect to $a$. From
 Eqs.~(\ref{eq12}) and (\ref{eq13}), we obtain
 \be{eq14}
 \frac{\dot{\alpha}}{\alpha}=\frac{Q}{6\rho_\Lambda}=-\frac{H}
 {6(1+f)}\left(a\frac{f^\prime}{f}-3\right)=-\frac{EH_0}
 {6(1+f)}\left(a\frac{f^\prime}{f}-3\right)\,,
 \ee
 where $E\equiv H/H_0$. If $f\propto a^3$, it is easy to see
 that $\dot{\alpha}=0$, namely $\alpha=const.$  On the other
 hand, one can recast the total energy conservation equation
 $\dot{\rho}_{tot}+3H\rho_{tot}(1+w_{tot})=0$ as
 \be{eq15}
 \frac{d\ln\rho_{tot}}{d\ln a} = -3\left(1-\Omega_\Lambda\right)\,,
 \ee
 in which we have used
 $w_{tot}=p_{tot}/\rho_{tot}=\Omega_\Lambda w_\Lambda+\Omega_m w_m$
 and $w_\Lambda=-1$, $w_m=0$. We can integrate Eq.~(\ref{eq15})
 to get
 \be{eq16}
 \rho_{tot}=\exp\left(\int_{const.}^a\left(-3+3\Omega_\Lambda
 \right)\frac{d\tilde{a}}{\tilde{a}}\right)=\rho_{tot,0\,}
 \exp\left(\int_1^a\left(-3+3\Omega_\Lambda
 \right)\frac{d\tilde{a}}{\tilde{a}}\right)\,,
 \ee
 where $const.$ is an integration constant. Using
 Eqs.~(\ref{eq12}), (\ref{eq16}) and $H^2=8\pi G\rho_{tot}/3$,
 we find
 \be{eq17}
 E^2\equiv\frac{H^2}{H_0^2}=a^{-3}\exp\left(
 \int_1^a \frac{3f}{1+f}\frac{d\tilde{a}}{\tilde{a}}\right)\,.
 \ee
 Noting $\rho_\Lambda\propto\alpha^{-6}$ and $\rho_\Lambda/
 \rho_{\Lambda 0}=\Omega_\Lambda E^2/\Omega_{\Lambda 0}=
 \Omega_\Lambda E^2/(1-\Omega_{m0})$, we have
 \be{eq18}
 \frac{\Delta\alpha}{\alpha}\equiv\frac{\alpha-\alpha_0}{\alpha_0}=
 \left(\frac{\Omega_\Lambda E^2}{1-\Omega_{m0}}\right)^{-1/6}-1 \,,
 \ee
 where $\Omega_\Lambda$ and $E^2$ are given in Eqs.~(\ref{eq12}) and
 (\ref{eq17}), respectively. It is easy to check that if
 $f\propto a^3$, we obtain $\Delta\alpha/\alpha=0$, namely
 $\alpha=const.$ In summary, if the function $f(a)$ is given,
 one can get the cosmic expansion history from Eq.~(\ref{eq17}), and
 the variation of the fine structure ``constant'' $\alpha$
 from Eqs.~(\ref{eq18}) or (\ref{eq14}). Finally, it is worth noting
 that by definition (\ref{eq11}), we obtain
 \be{eq19}
 f_0=f(a=1)=\frac{\rho_{\Lambda 0}}{\rho_{m0}}
 =\frac{1}{\Omega_{m0}}-1\,,
 \ee
 which is useful to fix one of the model parameters in $f(a)$.

%============================= section 2.2 ===================================

\subsection{Type II models}\label{sec2b}

Inspired by e.g.~\cite{r24,r26,r29}, the type II models are
 characterized by
 \be{eq20}
 \rho_m=\rho_{m0}\, a^{-3+\epsilon (a)}\,,
 \ee
 where $\epsilon(a)$ can be any function of scale factor $a$.
 Obviously, $\epsilon(a)\equiv 0$ corresponds to $\Lambda$CDM
 model whose $\rho_m=\rho_{m0}\, a^{-3}$. Substituting
 Eq.~(\ref{eq20}) into Eq.~(\ref{eq10}), we find that the
 corresponding interaction term is given by
 \be{eq21}
 Q=H\rho_m\left[\,\epsilon(a)+a\epsilon^\prime (a)\ln a\,\right]\,.
 \ee
 Substituting Eqs.~(\ref{eq20}) and (\ref{eq21})
 into Eq.~(\ref{eq9}), we obtain
 \be{eq22}
 \frac{d\rho_\Lambda}{da}=-\rho_{m0}\, a^{-4+\epsilon (a)}
 \left[\,\epsilon(a)+a\epsilon^\prime (a)\ln a\,\right]\,,
 \ee
 which can be integrated to get
 \be{eq23}
 \rho_\Lambda = \rho_{m0}\,\eta (a)+\rho_{\Lambda 0}\,,
 \ee
 where
 \be{eq24}
 \eta(a)\equiv\int_a^1 \tilde{a}^{-4+\epsilon(\tilde{a})}
 \left[\,\epsilon(\tilde{a})+\tilde{a}\epsilon^\prime (\tilde{a})\ln
 \tilde{a}\,\right]d\tilde{a}\,.
 \ee
 Substituting Eqs.~(\ref{eq20}) and (\ref{eq23}) into Friedmann
 equation $H^2=8\pi G(\rho_\Lambda+\rho_m)/3$, we find
 \be{eq25}
 E^2\equiv\frac{H^2}{H_0^2}=\Omega_{m0}\left[\, a^{-3+\epsilon(a)}+
 \eta(a)\,\right]+\left(1-\Omega_{m0}\right)\,.
 \ee
 Using Eqs.~(\ref{eq20}), (\ref{eq21}) and (\ref{eq23}),
 we have
 \be{eq26}
 \frac{\dot{\alpha}}{\alpha}=\frac{Q}{6\rho_\Lambda}=\frac{EH_0}{6}
 \frac{\Omega_{m0}\, a^{-3+\epsilon(a)}}{1+\Omega_{m0}\left[\,
 \eta(a)-1\,\right]}\left[\,\epsilon(a)+a\epsilon^\prime (a)\ln a\,
 \right]\,,
 \ee
 where $\eta(a)$ and $E$ are given in Eqs.~(\ref{eq24}) and
 (\ref{eq25}), respectively. On the other hand,
 using Eq.~(\ref{eq23}) and noting $\rho_\Lambda\propto\alpha^{-6}$,
 we obtain
 \be{eq27}
 \frac{\Delta\alpha}{\alpha}\equiv\frac{\alpha-\alpha_0}{\alpha_0}=
 \left[\,1+\frac{\Omega_{m0}\,\eta(a)}{1-\Omega_{m0}}\,
 \right]^{-1/6}-1\,,
 \ee
 where $\eta(a)$ is given in Eq.~(\ref{eq24}). So, if the
 function $\epsilon(a)$ is given, one can get the cosmic
 expansion history from Eq.~(\ref{eq25}), and the variation
 of the fine structure ``constant'' $\alpha$ from Eqs.~(\ref{eq27})
 or (\ref{eq26}). In particular, if $\epsilon(a)\equiv 0$, it
 is easy to check that $\dot{\alpha}=0$ and $\Delta\alpha/\alpha=0$,
 namely $\alpha=const.$

%============================= section 3 ===================================

\section{Observational constraints on the models}\label{sec3}

%============================= section 3.1 ===================================

\subsection{Observational data}\label{sec3a}

In the literature, there are two kinds of observational data
 concerning the variation of the fine structure ``constant'',
 namely the data of $\Delta\alpha/\alpha$ and the data of
 $\dot{\alpha}/\alpha$. To the best of our knowledge, most of
 the observational data are given in terms of $\Delta\alpha/\alpha$,
 and only a few of the observational data are given in terms
 of $\dot{\alpha}/\alpha$. On the other hand, comparing
 $\dot{\alpha}/\alpha$ in Eqs.~(\ref{eq14}) and (\ref{eq26})
 with $\Delta\alpha/\alpha$ in Eqs.~(\ref{eq18})
 and (\ref{eq27}), it is easy to see that the number of free
 parameters in $\dot{\alpha}/\alpha$ is always more than
 the one in $\Delta\alpha/\alpha$, namely one more free
 parameter $H_0$ is required in $\dot{\alpha}/\alpha$ while
 $\Delta\alpha/\alpha$ need not. Due to the above two reasons,
 we only consider the observational data of
 $\Delta\alpha/\alpha$ in the present work.

Here, we consider the observational $\Delta\alpha/\alpha$
 dataset given in~\cite{r30}, which consists of
 293 $\Delta\alpha/\alpha$ data from the absorption systems
 in the spectra of distant quasars. This sample includes 154
 quasar absorption systems from the Very Large Telescope (VLT)
 in Chile, and 141 quasar absorption systems from the Keck
 Observatory in Hawaii. The full numerical data of these 295
 quasar absorption systems are available in~\cite{r31} or
 \cite{r32}. According to~\cite{r30} and the
 instructions of~\cite{r31,r32}, two outliers (J194454+770552
 at $z_{abs}=2.8433$, and J000448-415728 at $z_{abs}=1.5419$)
 should be removed. Therefore, there are 293 usable data
 in the final dataset, over the absorption redshift range
 $0.2223\leq z_{abs}\leq 4.1798$. Note that all these 293
 $\Delta\alpha/\alpha$ data are of
 ${\cal O}(10^{-5})$. The $\chi^2$ from these
 293 $\Delta\alpha/\alpha$ data is given by
 \be{eq28}
 \chi^2_\alpha=\sum\limits_i \frac{\left[\,
 (\Delta\alpha/\alpha)_{{\rm th},i}-
 (\Delta\alpha/\alpha)_{{\rm obs},i}\,\right]^2}{\sigma_i^2}\,,
 \ee
 where $\sigma_i^2=\sigma^2_{{\rm stat},i}+\sigma^2_{{\rm rand},i}$
 (see Sec.~3.5.3 of~\cite{r30} and
 the instructions of~\cite{r31,r32} for the technical details
 of $\sigma_{\rm rand}$ and the error budget). In fact, we
 have tested our two types of models with these
 293 $\Delta\alpha/\alpha$ data, and found that
 these $\Delta\alpha/\alpha$ data can tightly constrain the
 model parameters in $f(a)$ or $\epsilon(a)$, but
 the constraints on the model parameter $\Omega_{m0}$ are too
 loose. Therefore, other cosmological observations, for
 instance, type Ia supernovae (SNIa), cosmic microwave
 background (CMB), and baryon acoustic oscillation (BAO),
 are required to properly constrain the model parameter
 $\Omega_{m0}$.

We further consider the Union2.1 SNIa dataset~\cite{r33}
 consisting of 580 data points, which are given in terms of
 the distance modulus $\mu_{\rm obs}(z_i)$. The theoretical
 distance modulus is defined by
 \be{eq29}
 \mu_{\rm th}(z_i)\equiv 5\log_{10} D_L(z_i)+\tilde{\mu}_0\,,
 \ee
 where $\tilde{\mu}_0\equiv 42.3841-5\log_{10}h$ ($h$ is the
 Hubble constant $H_0$ in units of $100\,{\rm km/s/Mpc}$), and
 \be{eq30}
 D_L(z)=(1+z)\int_0^z\frac{d\tilde{z}}{E(\tilde{z};{\bf p})}\,,
 \ee
 in which $E\equiv H/H_0$, and ${\bf p}$ denotes the model
 parameters. The $\chi^2$ from 580 Union2.1 SNIa is given by
 \be{eq31}
 \chi^2_\mu ({\bf p})=\sum\limits_i\frac{\left[\,
 \mu_{\rm obs}(z_i)-\mu_{\rm th}(z_i)\,\right]^2}
 {\sigma_{\mu_{\rm obs}}^2(z_i)}\,.
 \ee
 The parameter $\tilde{\mu}_0$ (equivalent to $H_0$) is a
 nuisance parameter, but it is independent of the data points.
 One can perform a uniform marginalization over
 $\tilde{\mu}_0$. However, there is an alternative way.
 Following~\cite{r34}, the minimization can be made by
 expanding $\chi^2_{\mu}$ in Eq.~(\ref{eq31}) with respect to
 $\tilde{\mu}_0$ as
 \be{eq32}
 \chi^2_{\mu}({\bf p})=\tilde{A}-2\tilde{\mu}_0\tilde{B}
 +\tilde{\mu}_0^2\tilde{C}\,,
 \ee
 where
 \bea
 &\disp\tilde{A}({\bf p})=\sum\limits_i\frac{\left[\,
 \mu_{\rm obs}(z_i)-\mu_{\rm th}(z_i;\tilde{\mu}_0=0,{\bf p})\,
 \right]^2}{\sigma_{\mu_{\rm obs}}^2(z_i)}\,,\nonumber\\[1mm]
 &\disp\tilde{B}({\bf p})=\sum\limits_i\frac{\mu_{\rm obs}(z_i)
 -\mu_{\rm th}(z_i;\tilde{\mu}_0=0,{\bf p})}
 {\sigma_{\mu_{\rm obs}}^2(z_i)}\,,~~~~~~~~~
 \tilde{C}=\sum\limits_i\frac{1}
 {\sigma_{\mu_{\rm obs}}^2(z_i)}\,.\nonumber
 \eea
 Eq.~(\ref{eq32}) has a minimum
 for $\tilde{\mu}_0=\tilde{B}/\tilde{C}$ at
 \be{eq33}
 \tilde{\chi}^2_{\mu}({\bf p})=
 \tilde{A}({\bf p})-\frac{\tilde{B}({\bf p})^2}{\tilde{C}}\,.
 \ee
 Since $\chi^2_{\mu,\,min}=\tilde{\chi}^2_{\mu,\,min}$ (up to a
 constant), we can instead minimize $\tilde{\chi}^2_{\mu}$
 which is independent of $\tilde{\mu}_0$.

Since using the full data of CMB and BAO to perform a global
 fitting consumes a large amount of computation time and power,
 we instead use the shift parameter $R$ from the observation
 of CMB, and the distance parameter $A$ from the measurement
 of BAO, which are model-independent and contain the main
 information of the observations of CMB and BAO~\cite{r35},
 respectively. The shift parameter $R$ of CMB is defined
 by~\cite{r36,r35}
 \be{eq34}
 R\equiv\Omega_{m0}^{1/2}\int_0^{z_\ast}
 \frac{d\tilde{z}}{E(\tilde{z})}\,,
 \ee
 where the redshift of recombination $z_\ast$ is determined to
 be $1089.90$ by the Planck 2015 data~\cite{r37}. On the other
 hand, the Planck 2015 data have also determined the observed
 value of shift parameter $R_{obs}$ to be
 $1.7382\pm 0.0088$~\cite{r38}. The $\chi^2$ from CMB is given
 by $\chi^2_R=(R-R_{obs})^2/\sigma_R^2$. The distance parameter
 $A$ of the measurement of the BAO peak in the distribution of
 SDSS luminous red galaxies~\cite{r39} is given by
 \be{eq35}
 A\equiv\Omega_{m0}^{1/2}\,E(z_b)^{-1/3}\left[\frac{1}{z_b}
 \int_0^{z_b}\frac{d\tilde{z}}{E(\tilde{z})}\right]^{2/3},
 \ee
 where $z_b=0.35$. In~\cite{r39}, the value of $A$ has been
 determined to be $0.469\,(n_s/0.98)^{-0.35}\pm 0.017$. Here
 the scalar spectral index $n_s$ is taken to be $0.9741$ by
 the Planck 2015 data~\cite{r38}. The $\chi^2$ from BAO is
 given by $\chi^2_A=(A-A_{obs})^2/\sigma_A^2$.

The total $\chi^2$ from the combined $\Delta\alpha/\alpha$,
 SNIa, CMB and BAO data is given by
 \be{eq36}
 \chi^2=\chi^2_\alpha+\tilde{\chi}^2_\mu+\chi^2_R+\chi^2_A\,.
 \ee
 The best-fit model parameters are determined by minimizing
 the total $\chi^2$. As in~\cite{r34,r40}, the $68.3\%$
 confidence level is determined by
 $\Delta\chi^2\equiv\chi^2-\chi^2_{min}\leq 1.0$, $2.3$,
 $3.53$, $4.72$ for $n_p=1$, $2$, $3$, $4$, respectively,
 where $n_p$ is the number of free model parameters. Similarly,
 the $95.4\%$ confidence level is determined by
 $\Delta\chi^2\equiv\chi^2-\chi^2_{min}\leq 4.0$, $6.18$,
 $8.02$, $9.72$ for $n_p=1$, $2$, $3$, $4$, respectively.

\newpage  % used here just for a comfortable typesetting

%============================= Fig. 1 =================================

 \begin{center}
 \begin{figure}[tb]
 \centering
 \vspace{-4mm}  % used here just for a comfortable typesetting
 \includegraphics[width=0.45\textwidth]{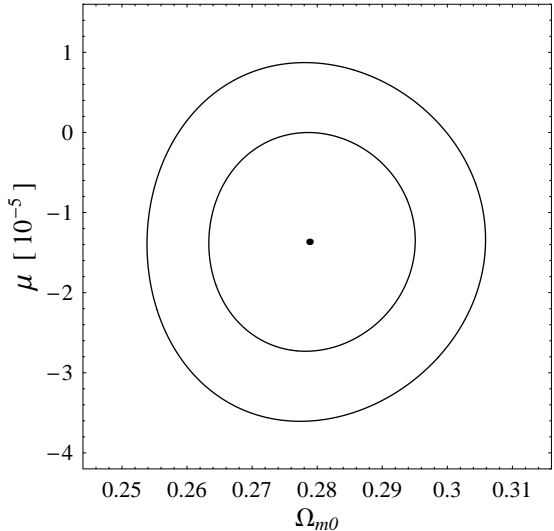}
 \caption{\label{fig1} The $68.3\%$ and $95.4\%$ confidence
 level contours in the $\Omega_{m0}-\mu$ plane for the type I
 model characterized by $f(a)=f_0\,a^\xi$ given in Eq.~(\ref{eq37}).
 The best-fit parameters are also indicated by the black solid
 point. Note that $\mu=\xi-3$ is given in units
 of $10^{-5}$. See the text for details.}
 \end{figure}
 \end{center}

%======================================================================

%============================= Fig. 2 =================================

 \begin{center}
 \begin{figure}[tb]
 \centering
 \vspace{-9mm}  % used here just for a comfortable typesetting
 \includegraphics[width=1.0\textwidth]{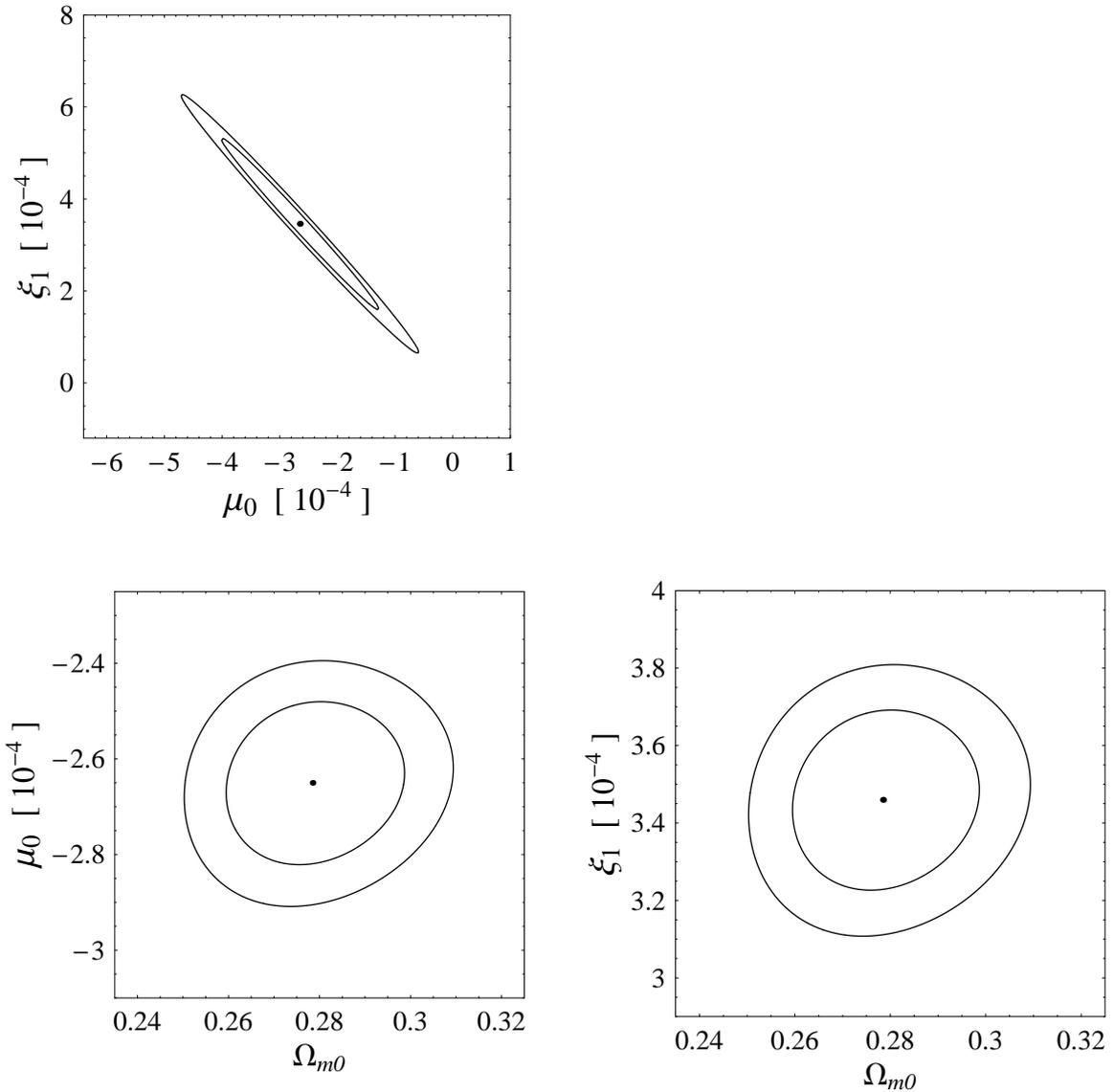}
 \caption{\label{fig2} The $68.3\%$ and $95.4\%$ confidence
 level contours in the $\mu_0-\xi_1$, $\Omega_{m0}-\mu_0$ and
 $\Omega_{m0}-\xi_1$ planes for the type I model characterized
 by $f(a)=f_0\,a^{\xi_0+\xi_1(1-a)}$ given in Eq.~(\ref{eq39}).
 The best-fit parameters are also indicated by the black solid
 points. Note that $\mu_0=\xi_0-3$ and $\xi_1$ are given in
 units of $10^{-4}$. See the text for details.}
 \end{figure}
 \end{center}

%======================================================================

\vspace{-24mm}  % used here just for a comfortable typesetting

%============================= section 3.2 ===================================

\subsection{Observational constraints on type I models}\label{sec3b}

Now, we consider the observational constraints on type I
 models introduced in Sec.~\ref{sec2a}. At first, we choose
 the simplest form of the function $f(a)$, namely
 \be{eq37}
 f(a)=f_0\, a^\xi\,,
 \ee
 where $\xi=const.$ and $f_0$ is given in Eq.~(\ref{eq19}).
 In this case, it is easy to find the explicit formula of
 $E^2$ in Eq.~(\ref{eq17}), namely
 \be{eq38}
 E^2=a^{-3}\left[\,\Omega_{m0}+\left(1-\Omega_{m0}\right)a^\xi
 \,\right]^{3/\xi}=(1+z)^3 \left[\,\Omega_{m0}
 +\left(1-\Omega_{m0}\right)(1+z)^{-\xi}\,\right]^{3/\xi}\,.
 \ee
 If $\xi=3$, it reduces to $\Lambda$CDM model and
 $\alpha=const.$ (nb. Eqs.~(\ref{eq14}) and (\ref{eq18})). It
 is natural to expect $\xi$ will be very close to $3$, since
 all the 293 $\Delta\alpha/\alpha$ data given
 in~\cite{r30,r31,r32} are of ${\cal O}(10^{-5})$. So, it is
 convenient to introduce $\mu=\xi-3$, and then we
 recast $\xi=\mu+3$. There are two free model parameters
 $\Omega_{m0}$ and $\mu$ (which is equivalent to $\xi$). By
 minimizing the corresponding total $\chi^2$ in
 Eq.~(\ref{eq36}), we find the best-fit model parameters
 $\Omega_{m0}=0.279$ and $\mu=\xi-3=-1.366\times 10^{-5}$,
 while $\chi^2_{min}=868.149$ and $\chi^2_{min}/dof=0.994$.
 In Fig.~\ref{fig1}, we also present the corresponding
 $68.3\%$ and $95.4\%$ confidence level contours in the
 $\Omega_{m0}-\mu$ plane. The parameter $\mu=\xi-3$ is tightly
 constrained to a narrow range of ${\cal O}(10^{-5})$, thanks
 to the 293 $\Delta\alpha/\alpha$ data of ${\cal O}(10^{-5})$.
 From Fig.~\ref{fig1}, we note that $\mu=\xi-3=0$
 (corresponding to $\Lambda$CDM model and $\alpha=const.$)
 deviates from the best fit beyond $1\sigma$, although it is
 still consistent with the data in $2\sigma$ region. Thus, the
 varying $\Lambda$ and $\alpha$ are
 favored by the observational data.

Next, we can generalize the simplest model in Eq.~(\ref{eq37})
 by allowing $\xi=\xi(a)$ is not a constant. Similar to the
 well-known Chevallier-Polarski-Linder (CPL) EoS
 parameterization $w=w_0+w_a(1-a)$~\cite{r41}, the simplest
 form for $\xi(a)$ is CPL-like,
 namely $\xi(a)=\xi_0+\xi_1(1-a)$. Noting that the Taylor
 series expansion of any (even unknown) function $F(x)$ is
 given by $F(x)=F(x_0)+F_1\,(x-x_0)+(F_2/\,2!)\,(x-x_0)^2+
 (F_3/\,3!)\,(x-x_0)^3+\dots$, the CPL-like
 $\xi(a)=\xi_0+\xi_1(1-a)$ can be regarded as the Taylor series
 expansion of $\xi(a)$ with respect to scale factor $a$ up to
 first order (linear expansion). Thus, it is well motivated to
 consider another type I model characterized by
 \be{eq39}
 f(a)=f_0\, a^{\xi(a)}\,,~~~~~~
 {\rm and}~~~~~~\xi(a)=\xi_0+\xi_1(1-a)\,,
 \ee
 where $\xi_0$, $\xi_1$ are constants, and $f_0$ is given in
 Eq.~(\ref{eq19}). If $\xi_0=3$ and $\xi_1=0$, it reduces to
 $\Lambda$CDM model and $\alpha=const.$ (nb.~Eqs.~(\ref{eq14})
 and (\ref{eq18})). It is natural to expect $\xi_0$ will be
 very close to $3$, since all the 293 $\Delta\alpha/\alpha$
 data given in~\cite{r30,r31,r32} are of ${\cal O}(10^{-5})$.
 So, it is convenient to introduce $\mu_0=\xi_0-3$, and then
 we recast $\xi_0=\mu_0+3$. There are three free model
 parameters $\Omega_{m0}$, $\xi_1$ and $\mu_0$ (which is equivalent
 to~$\xi_0$). Note that there is no analytical formula for
 $E^2$ in this case, but we can get it by using numerical
 integration in Eq.~(\ref{eq17}). By minimizing
 the corresponding total $\chi^2$ in Eq.~(\ref{eq36}), we find
 the best-fit model parameters
 $\Omega_{m0}=0.278$, $\mu_0=\xi_0-3=-2.650\times 10^{-4}$,
 and $\xi_1=3.460\times 10^{-4}$, while $\chi^2_{min}=856.005$
 and $\chi^2_{min}/dof=0.982$. Note that this $\chi^2_{min}$
 is significantly smaller than the one of the model with
 Eq.~(\ref{eq37}), namely $856.005$ vs.~$868.149$, just at the
 price of adding only one free model parameter. In Fig.~\ref{fig2},
 we also present the corresponding $68.3\%$ and $95.4\%$ confidence
 level contours in the $\mu_0-\xi_1$, $\Omega_{m0}-\mu_0$ and
 $\Omega_{m0}-\xi_1$ planes. Both the parameters $\mu_0=\xi_0-3$ and
 $\xi_1$ are tightly constrained to the narrow ranges of
 ${\cal O}(10^{-4})$, thanks to the 293 $\Delta\alpha/\alpha$
 data of ${\cal O}(10^{-5})$. From Fig.~\ref{fig2}, we note that
 $\mu_0=\xi_0-3=0$ and $\xi_1=0$ (corresponding to $\Lambda$CDM
 model and $\alpha=const.$) deviate from the best
 fit far beyond $2\sigma$. Thus, the varying
 $\Lambda$ and $\alpha$ are favored by the observational data.

%============================= Fig. 3 =================================

 \begin{center}
 \begin{figure}[tb]
 \centering
 \vspace{-4mm}  % used here just for a comfortable typesetting
 \includegraphics[width=0.45\textwidth]{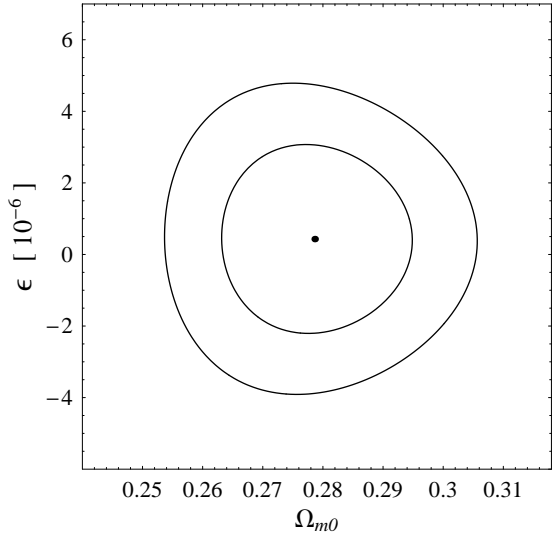}
 \caption{\label{fig3} The $68.3\%$ and $95.4\%$ confidence
 level contours in the $\Omega_{m0}-\epsilon$ plane for the
 type II model characterized by $\epsilon=const.$ given in
 Eq.~(\ref{eq40}). The best-fit parameters are also indicated
 by the black solid point. Note that $\epsilon$ is given in
 units of $10^{-6}$. See the text for details.}
 \end{figure}
 \end{center}

%======================================================================

%============================= Fig. 4 =================================

 \begin{center}
 \begin{figure}[tb]
 \centering
 \vspace{-9mm}  % used here just for a comfortable typesetting
 \includegraphics[width=1.0\textwidth]{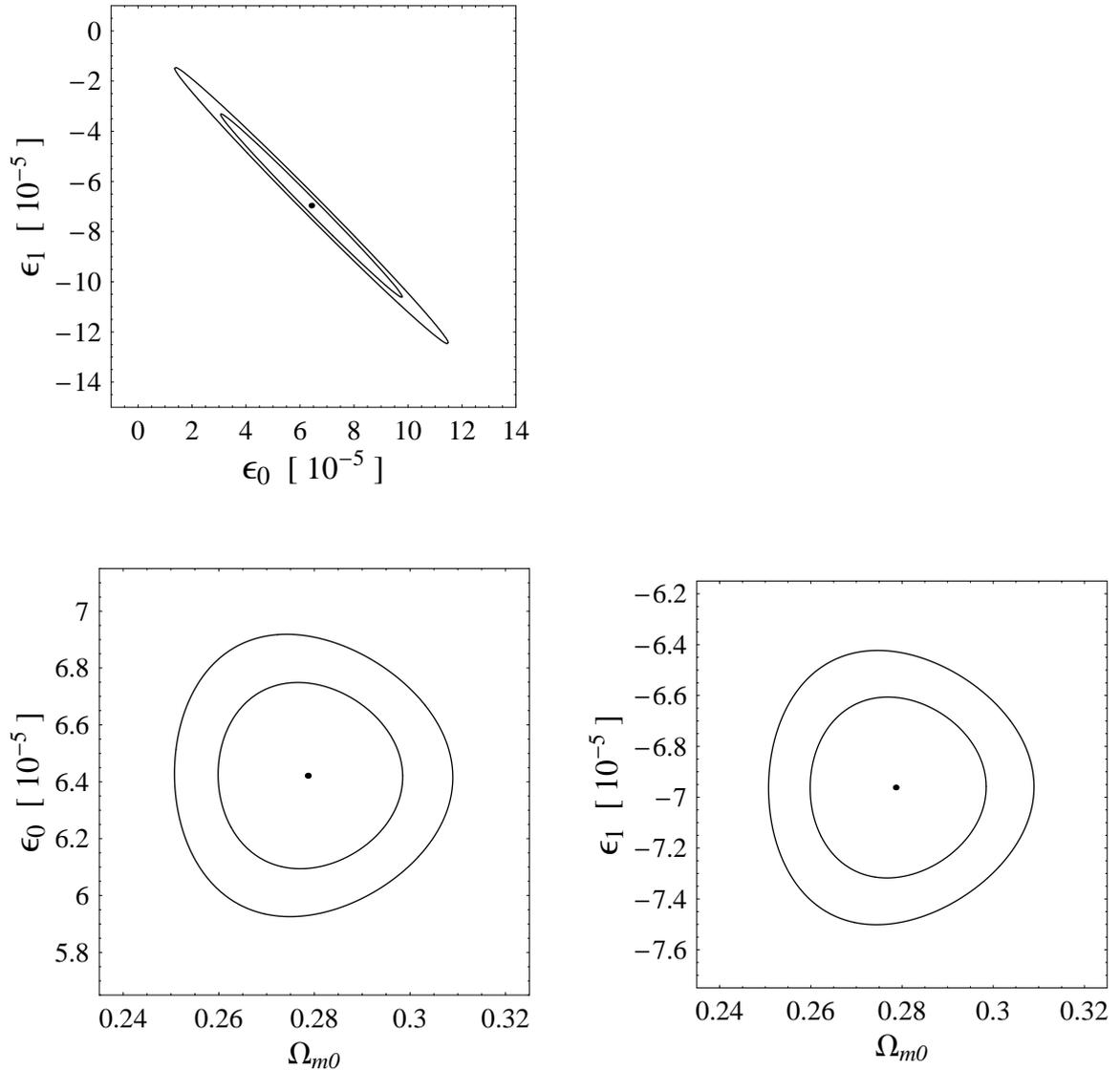}
 \caption{\label{fig4} The $68.3\%$ and $95.4\%$ confidence
 level contours in the $\epsilon_0-\epsilon_1$,
 $\Omega_{m0}-\epsilon_0$ and $\Omega_{m0}-\epsilon_1$
 planes for the type II model characterized by
 $\epsilon(a)=\epsilon_0+\epsilon_1(1-a)$ given in Eq.~(\ref{eq43}).
 The best-fit parameters are also indicated by the black solid
 points. Note that $\epsilon_0$ and $\epsilon_1$ are given in
 units of $10^{-5}$. See the text for details.}
 \end{figure}
 \end{center}

%======================================================================

\vspace{-24mm}  % used here just for a comfortable typesetting

%============================= section 3.3 ===================================

\subsection{Observational constraints on type II models}\label{sec3c}

Let us turn to the observational constraints on type II
 models introduced in Sec.~\ref{sec2b}. Obviously, the
 simplest type II model is given by
 \be{eq40}
 \rho_m=\rho_{m0}\, a^{-3+\epsilon}\,,
 \ee
 where $\epsilon\not=3$ is a constant. If $\epsilon=0$, it
 reduces to $\Lambda$CDM model and $\alpha=const.$ (nb.
 Eqs.~(\ref{eq26}) and (\ref{eq27})). For $\epsilon(a)=\epsilon
 \not=3$, we find the analytical formulas for $\eta(a)$
 and $E^2$ in Eqs.~(\ref{eq24}) and (\ref{eq25}), namely
 \bea
 &\disp\eta=\frac{\epsilon}{\epsilon-3}\left(1-a^{-3+\epsilon}
 \right)=\frac{\epsilon}{\epsilon-3}\left[\,1-(1+z)^{3-
 \epsilon}\,\right]\,,\label{eq41}\\[1.5mm]
 &\disp E^2=\frac{\Omega_{m0}}{\epsilon-3}\left(
 \epsilon-3a^{-3+\epsilon}\right)+\left(1-\Omega_{m0}\right)=
 \frac{\Omega_{m0}}{\epsilon-3}\left[\,\epsilon-3(1+z)^{3-
 \epsilon}\,\right]+\left(1-\Omega_{m0}\right)\,.\label{eq42}
 \eea
 There are two free model parameters, namely $\Omega_{m0}$ and
 $\epsilon$. By minimizing the corresponding total $\chi^2$ in
 Eq.~(\ref{eq36}), we find the best-fit model parameters
 $\Omega_{m0}=0.279$ and $\epsilon=0.430\times 10^{-6}$,
 while $\chi^2_{min}=870.391$ and $\chi^2_{min}/dof=0.997$.
 In Fig.~\ref{fig3}, we also present the corresponding
 $68.3\%$ and $95.4\%$ confidence level contours in the
 $\Omega_{m0}-\epsilon$ plane. The parameter $\epsilon$ is
 tightly constrained to a narrow range of ${\cal O}(10^{-6})$,
 thanks to the 293 $\Delta\alpha/\alpha$ data of
 ${\cal O}(10^{-5})$. From Fig.~\ref{fig3}, we see that
 $\epsilon=0$ (corresponding to $\Lambda$CDM model and
 $\alpha=const.$) is fully consistent with the observational
 data (in fact it is close to the best fit). So, $\Lambda$ and
 $\alpha$ can be non-varying in the type II model characterized
 by Eq.~(\ref{eq40}).

Next, we consider another type II model characterized by a
 CPL-like $\epsilon(a)$, namely
 \be{eq43}
 \epsilon(a)=\epsilon_0+\epsilon_1(1-a)\,,
 \ee
 where $\epsilon_0$ and $\epsilon_1$ are constants. If
 $\epsilon_0=\epsilon_1=0$, it reduces to $\Lambda$CDM model
 and $\alpha=const.$ (nb. Eqs.~(\ref{eq26}) and (\ref{eq27})).
 As mentioned above, this CPL-like
 $\epsilon(a)=\epsilon_0+\epsilon_1(1-a)$ can be regarded as
 the Taylor series expansion of $\epsilon(a)$ with respect to
 the scale factor $a$ up to first order (linear expansion), and
 hence it is well motivated. There are three free model parameters,
 namely $\Omega_{m0}$, $\epsilon_0$ and $\epsilon_1$. Note
 that there are~no analytical formulas for $\eta(a)$ and $E^2$
 in this case, but we can get $\eta(a)$ by using numerical
 integration in Eq.~(\ref{eq24}) and then $E^2$
 in Eq.~(\ref{eq25}) is ready. By minimizing the corresponding
 total $\chi^2$ in Eq.~(\ref{eq36}), we find the best-fit model
 parameters $\Omega_{m0}=0.279$,
 $\epsilon_0=6.421\times 10^{-5}$
 and $\epsilon_1=-6.962\times 10^{-5}$, while
 $\chi^2_{min}=857.605$ and $\chi^2_{min}/dof=0.983$. Note that
 this $\chi^2_{min}$ is significantly smaller than the
 one of the model with Eq.~(\ref{eq40}), namely $857.605$
 vs.~$870.391$, just at the price of adding only one free model
 parameter. In Fig.~\ref{fig4}, we also present
 the corresponding $68.3\%$ and $95.4\%$ confidence level
 contours in the $\epsilon_0-\epsilon_1$,
 $\Omega_{m0}-\epsilon_0$ and $\Omega_{m0}-\epsilon_1$
 planes. Both the parameters $\epsilon_0$ and
 $\epsilon_1$ are tightly constrained to the narrow ranges of
 ${\cal O}(10^{-5})$, thanks to the 293 $\Delta\alpha/\alpha$
 data of ${\cal O}(10^{-5})$. From Fig.~\ref{fig4}, we note
 that $\epsilon_0=\epsilon_1=0$ (corresponding to $\Lambda$CDM
 model and $\alpha=const.$) deviate from the best
 fit far beyond $2\sigma$. This indicates that the varying
 $\Lambda$ and $\alpha$ are favored by the observational data.

%============================= Fig. 5 =================================

 \begin{center}
 \begin{figure}[tb]
 \centering
 \vspace{-9mm}  % used here just for a comfortable typesetting
 \includegraphics[width=0.96\textwidth]{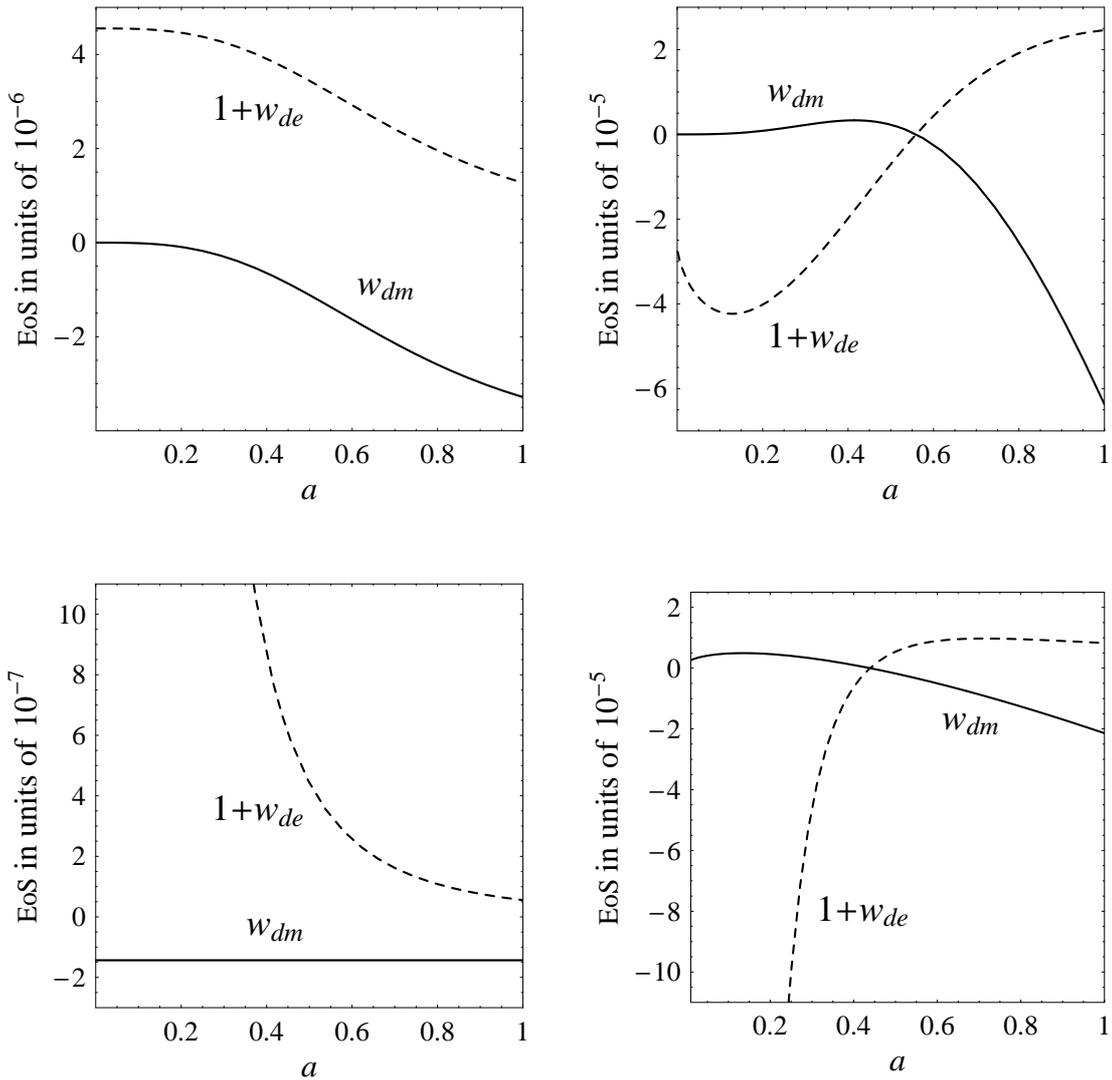}
 \caption{\label{fig5} The effective EoS of ``warm dark
 matter'' and ``dark energy'' given in terms of $w_{dm}$
 (solid lines) and $1+w_{de}$ (dashed lines), for (top-left
 panel) the type I model with $f(a)=f_0\,a^\xi$
 in Eq.~(\ref{eq37}), (top-right panel) the type I model with
 $f(a)=f_0\,a^{\xi_0+\xi_1(1-a)}$ in Eq.~(\ref{eq39}),
 (bottom-left panel) the type II model with $\epsilon=const.$
 in Eq.~(\ref{eq40}), (bottom-right panel) the type II model
 with $\epsilon(a)=\epsilon_0+\epsilon_1(1-a)$ in
 Eq.~(\ref{eq43}), while the corresponding best-fit model
 parameters obtained in Sec.~\ref{sec3} are taken. Note that
 they are given in units of $10^{-5}$, $10^{-6}$
 or $10^{-7}$. See the text for details.}
 \end{figure}
 \end{center}

%======================================================================

\vspace{-11mm}  % used here just for a comfortable typesetting

%============================= section 4 ===================================

\section{Varying cosmological constant and warm dark matter}\label{sec4}

In the previous sections, we turned the varying cosmological
 constant model into an interacting vacuum energy model. The
 vacuum energy interacts with the pressureless matter by
 exchanging energy between them. In this section, we would
 like to view this model from another perspective. As is shown
 in e.g.~\cite{r42}, an interacting dark energy model can be
 equivalent to a warm dark matter model without interaction
 between dark energy and dark matter, while these two different
 kinds of models can share both the same cosmic expansion
 history and growth history. To keep things simple, here we
 only consider the models from the side of expansion history.

Although the cold dark matter (CDM) model is very successful
 in many fields, it has been seriously challenged recently.
 We refer to e.g.~\cite{r43} for the detailed reviews on these
 challenges. Recently, warm dark matter (WDM) remarkably rose
 as an alternative of CDM. We refer to e.g.~\cite{r44} for
 several comprehensive reviews. The leading WDM candidate is
 the keV scale sterile neutrino. In fact, the keV scale WDM is
 an intermediate case between the eV scale hot dark matter
 (HDM) and the GeV scale CDM. Unlike CDM which is challenged
 on the small/galactic scale, it is claimed that WDM can
 successfully reproduce the astronomical observations over
 all the scales (from small/galactic to large/cosmological
 scales)~\cite{r44}. One of the key differences between WDM
 and CDM is their EoS. WDM has a fairly small but non-zero
 EoS, while the EoS of CDM is exactly zero. In the literature,
 many attempts have been made to determine the EoS of dark
 matter (see e.g.~\cite{r45,r46,r47,r48,r49}), and it is found that
 the EoS of WDM are of ${\cal O}(10^{-6})$, ${\cal O}(10^{-5})$
 or ${\cal O}(10^{-3})$ (depending on the working
 assumptions and the observational data in use).

Let us come back to the starting point Eqs.~(\ref{eq9}) and
 (\ref{eq10}), and view them from another perspective. In
 the form of Eqs.~(\ref{eq9}) and (\ref{eq10}), the vacuum
 energy (whose EoS is $w_\Lambda=-1$) interacts with the
 cold dark matter (whose EoS is $w_m=0$) through an interaction
 $Q=6\rho_\Lambda\dot{\alpha}/\alpha\not=0$. Now, we recast them as
 \be{eq44}
 \dot{\rho}_\Lambda+3H\rho_\Lambda\left(1+
 w_\Lambda^{\rm eff}\right)=0\,,~~~~~~~~~~
 \dot{\rho}_m+3H\rho_m\left(1+w_m^{\rm eff}\right)=0\,,
 \ee
 where
 \be{eq45}
 w_{dm}\equiv w_m^{\rm eff}=-\frac{Q}{3H\rho_m}\,,~~~~~~~~~~
 1+w_{de}\equiv 1+w_\Lambda^{\rm eff} = \frac{Q}{3H \rho_\Lambda}\,.
 \ee
 In this new form, the varying cosmological constant model becomes a
 model containing ``dark energy'' (whose EoS $w_{de}\not=-1$)
 and ``warm dark matter'' (whose EoS $w_{dm}\not=0$), but there
 is no interaction between them. Since the observational data
 concerning the time variation of $\alpha$
 are of ${\cal O}(10^{-5})$, it is natural to expect that
 $w_{dm}\sim 1+w_{de}\sim{\cal O}(10^{-5})$ or smaller.

For type I models introduced in Sec.~\ref{sec2a}, substituting
 Eq.~(\ref{eq13}) into Eq.~(\ref{eq45}), we have
 \be{eq46}
 w_{dm}=\frac{\Omega_\Lambda}{3}\left(a\frac{f^\prime}{f}
 -3\right)\,,~~~~~~~
 1+w_{de}= -\frac{\Omega_m}{3}\left(a\frac{f^\prime}{f}-3\right)\,,
 \ee
 where $\Omega_\Lambda$ and $\Omega_m$ are given
 in Eq.~(\ref{eq12}). Thus, it is easy to find the evolutions
 of $w_{dm}$ and $w_{de}$ if $f(a)$ is given. In the top panels
 of Fig.~\ref{fig5}, we plot $w_{dm}$ and $1+w_{de}$ as
 functions of the scale factor $a$ for the type I models with
 $f(a)=f_0\,a^\xi$ in Eq.~(\ref{eq37}) and
 $f(a)=f_0\,a^{\xi_0+\xi_1(1-a)}$ in Eq.~(\ref{eq39}), while
 the corresponding best-fit model parameters obtained in
 Sec.~\ref{sec3b} are taken. As expected above, they are of
 order $10^{-6}$ or $10^{-5}$. Thus, the effective ``warm dark
 matter'' from type I models of the varying cosmological
 constant $\Lambda\propto\alpha^{-6}$ is fully consistent with
 the observational constraints
 on WDM (e.g.~\cite{r45,r46,r47,r48,r49,r58}).

For type II models introduced in Sec.~\ref{sec2b}, substituting
 Eqs.~(\ref{eq21}), (\ref{eq20}) and (\ref{eq23})
 into Eq.~(\ref{eq45}), we get
 \bea
 &\disp w_{dm}=-\frac{1}{3}\left[\,\epsilon(a)+
 a\epsilon^\prime (a)\ln a\,\right]\,,\label{eq47}\\[1.5mm]
 &\disp 1+w_{de}=\frac{\Omega_{m0}}{3}\frac{a^{-3+\epsilon(a)}}
 {1+\Omega_{m0}\left[\,\eta(a)-1\,\right]}\left[\,\epsilon
 (a)+a\epsilon^\prime (a)\ln a\,\right]\,,\label{eq48}
 \eea
 where $\eta(a)$ is given in Eq.~(\ref{eq24}). Thus, it is
 easy to find the evolutions of $w_{dm}$ and $w_{de}$ if
 $\epsilon(a)$ is given. In the case
 of $\epsilon(a)=\epsilon=const.$, there is an explicit
 formula in Eq.~(\ref{eq41}) for $\eta(a)$. However, in the
 case of $\epsilon(a)=\epsilon_0+\epsilon_1(1-a)$, we should
 get $\eta(a)$ by using numerical integration
 in Eq.~(\ref{eq24}). In the bottom panels of Fig.~\ref{fig5},
 we plot $w_{dm}$ and $1+w_{de}$ as functions of the scale
 factor $a$ for the type II models with $\epsilon=const.$ in
 Eq.~(\ref{eq40}) and $\epsilon(a)=\epsilon_0+\epsilon_1(1-a)$
 in Eq.~(\ref{eq43}), while the corresponding best-fit model
 parameters obtained in Sec.~\ref{sec3c} are taken. Clearly,
 they are of order $10^{-5}$ or $10^{-7}$. Again, the effective
 ``warm dark matter'' from type II models of the varying
 cosmological constant $\Lambda\propto\alpha^{-6}$ is also
 fully consistent with the observational constraints on
 WDM (e.g.~\cite{r45,r46,r47,r48,r49,r58}).

%============================= section 5 ===================================

\section{Concluding remarks}\label{sec5}

In this work, we considered the cosmological constant model
 $\Lambda\propto\alpha^{-6}$, which is well motivated from
 three independent approaches as mentioned in
 Sec.~\ref{sec1}. As is well known, in the passed 18 years,
 the hint of varying fine structure constant $\alpha$ was
 found, and the observational data of varying $\alpha$
 were accumulated. Nowadays, a time-varying $\alpha$ has
 been extensively discussed in the community. If
 $\Lambda\propto\alpha^{-6}$ is right, it means that the
 cosmological constant $\Lambda$ should also be varying. In
 this work, we tried to develop a suitable framework to model
 this varying cosmological constant
 $\Lambda\propto\alpha^{-6}$, in which we view it from an
 interacting vacuum energy perspective. We proposed two types
 of models to describe the evolutions of $\Lambda$ and
 $\alpha$. Then, we considered the observational constraints
 on these models, by using the 293 $\Delta\alpha/\alpha$ data
 from the absorption systems in the spectra of distant quasars,
 and the data of SNIa, CMB, BAO. We found that the model
 parameters can be tightly constrained to the narrow ranges of
 ${\cal O}(10^{-5})$ typically, thanks to the 293
 $\Delta\alpha/\alpha$ observational data of
 ${\cal O}(10^{-5})$. In particular, 3 of 4 models considered
 in this work favor the varying $\Lambda$ and $\alpha$, while
 $\Lambda$CDM model and $\alpha=const.$ deviate from the best
 fit beyond $2\sigma$ or at least $1\sigma$. On the other hand,
 we can also view the varying cosmological constant model
 $\Lambda\propto\alpha^{-6}$ from another perspective, namely
 it can be equivalent to a model containing ``dark energy''
 (whose EoS $w_{de}\not=-1$) and ``warm dark matter'' (whose
 EoS $w_{dm}\not=0$), but there is no interaction between them.
 We derived the effective EoS of ``warm dark matter'' and
 ``dark energy'', and found that they are fully consistent
 with the observational constraints on warm dark matter. In
 summary, we consider that the varying cosmological constant
 model $\Lambda\propto\alpha^{-6}$ is viable and deserves
 further studies.

Some remarks are in order. First, although the cosmological
 constant $\Lambda\propto\alpha^{-6}$ is derived from three
 independent approaches as mentioned in Sec.~\ref{sec1} (see
 also e.g.~\cite{r59}), the underlying fundamental theory for
 it is still unknown. Varying fundamental constants require
 new physics~\cite{r21}. It is commonly believed
 that the cosmological constant problem can only be solved
 ultimately in a unified theory of quantum
 gravity and the standard model of electroweak and strong
 interactions, which is still absent so far. Nevertheless,
 we consider that the studies on such a cosmological constant
 $\Lambda\propto\alpha^{-6}$ might shed new light
 on the possible ways to the unknown underlying theory.

Second, we have tightly constrained the model parameters
 besides $\Omega_{m0}$, namely $\xi$, $\xi_0$, $\xi_1$, $\epsilon$,
 $\epsilon_0$ and $\epsilon_1$, to the narrow ranges of
 ${\cal O}(10^{-5})$ typically, mainly by using the 293
 $\Delta\alpha/\alpha$ observational data from the absorption
 systems in the spectra of distant quasars. In fact, these
 parameters were confronted with only the observations of
 SNIa, CMB and BAO in e.g.~\cite{r29}, and the corresponding
 constraints are of ${\cal O}(1)$. So, the significant leap
 from ${\cal O}(1)$ to ${\cal O}(10^{-5})$ shows the great
 power of the 293 $\Delta\alpha/\alpha$ observational data.
 In fact, these 293 $\Delta\alpha/\alpha$ data have been used
 in many issues (see e.g.~\cite{r50,r51}). We advocate the
 further uses of these $\Delta\alpha/\alpha$ observational data
 in relevant studies.

Third, in addition to the well-known evidence of the time
 variation in the fine structure constant $\alpha$, it
 was claimed that $\alpha$ is also spatially
 varying~\cite{r30,r52}. If $\Lambda\propto\alpha^{-6}$ is
 right, the cosmological constant should be not only time-dependent
 but also space-dependent. This might bring about
 new features to this field, and deserve further detailed
 studies. For example, it is claimed that there exists a
 preferred direction in the CMB temperature map (known as
 the ``Axis of Evil'' in the literature)~\cite{r53}, the
 distribution of SNIa or gamma-ray bursts~\cite{r54,r55,r56},
 and the quasar optical polarization data~\cite{r57}. If the
 cosmological constant $\Lambda\propto\alpha^{-6}$ is also
 space-dependent, it might be responsible for the possible
 anisotropy in the (accelerated) expansion of the
 universe. We leave this interesting issue to future work.

Fourth, besides the 293 $\Delta\alpha/\alpha$ data from the
 absorption systems in the spectra of distant quasars, there
 are more $\Delta\alpha/\alpha$ observational data from other
 types of observations, for examples, atomic clocks, Oklo
 natural nuclear reactor, meteorite dating, CMB, big bang
 nucleosynthesis. We refer to e.g.~\cite{r21} for
 a comprehensive review. Thus, it is interesting to consider
 the constraints from these observational data. Since these
 data are subtle in some sense~\cite{r21,r23}, we also leave
 this to future work.

Fifth, let us turn to the varying cosmological constant model
 itself. In the present work, we considered four particular
 parameterizations of the functions $f(a)$ and $\epsilon(a)$.
 In fact, one can instead consider other parameterizations, for
 instance, $f(a)$, $\xi(a)$ or $\epsilon(a)$ characterized by
 $c_0+c_1\ln a$ or $c_0 a^{c_1}$~\cite{r29}. On the other
 hand, in the present work we assumed that only the fine
 structure constant $\alpha$ is varying, and the other
 fundamental constants $G$, $c$, $\hbar$, $m_e$ do not vary
 indeed. However, the varying $G$, $c$, $\hbar$, $m_e$ models
 do exist in the literature (see e.g.~\cite{r21} for a
 comprehensive review). Since the cosmological constant
 $\Lambda$ given in Eqs.~(\ref{eq1}) or (\ref{eq2}) also
 depends on $G$, $c$, $\hbar$ and $m_e$, there are diverse
 variants of the varying cosmological constant model in fact.
 These variants might bring about new features. Since this
 is beyond the scope of the present work, we again leave
 it to future work.

Sixth, it is worth noting that the data analysis based on
 Eq.~(\ref{eq36}) can only give a rough indication and cannot
 be used to infer that any of the models considered in the
 present work is better than $\Lambda$CDM (we
 thank Prof.~Dominik Schwarz for pointing out this issue),
 because (a) the SNIa light curve fitters assumed
 $\Lambda$CDM. So, if one wants to fit a model other than
 $\Lambda$CDM, the light curve fitting procedure should be
 redone within the new model. (b) the use of the parameters
 $R$ and $A$ from CMB and BAO measurements is crude, since
 they cannot take some complicated effects (e.g. the integrated
 Sachs-Wolfe effect) into account. (c) Eq.~(\ref{eq36})
 implicitly assumes that $\Delta\alpha/\alpha$ data, SNIa,
 CMB and BAO have equal statistical weight, but this is
 questionable. In addition, a full Markov Chain Monte Carlo
 (MCMC) analysis of the CMB data should be used to further
 test the models considered in this work (we
 thank Prof.~Dominik Schwarz for pointing out this issue),
 and we leave it to future work.

Finally, it is important to clarify the two different (but
 equivalent) perspectives on the varying cosmological constant
 model considered in this work. The first perspective is to
 regard the varying cosmological constant as a fluid {\em not}
 interacting with dark matter. In fact, this is the case
 considered in Sec.~\ref{sec4}. The conservation equation of
 this fluid is given by Eq.~(\ref{eq44}), in the form of
 $\dot{\rho}_{de}+3H\rho_{de}\left(1+w_{de}\right)=0$. In
 this case, we stress that the EoS of this fluid is not
 $w_{de}=-1$. As is clearly shown in e.g. Fig.~\ref{fig5}, the
 EoS of this fluid is time-dependent, rather than constant.
 Indeed, the EoS of this fluid $w_{de}\not=-1$ in this case.
 So, one cannot say $\rho_{de}=const.$ We have not assumed
 $w_{de}=-1$ in this perspective indeed, and we refer to
 Sec.~\ref{sec4} for detailed discussions. On the other hand,
 the second perspective on the varying cosmological constant
 model is considered in Secs.~\ref{sec2} and \ref{sec3}. In
 this case, one can regard the varying cosmological constant
 as a fluid interacting with dark matter (different from the
 first perspective considered in Sec.~\ref{sec4}). Now, the
 conservation equation of this fluid is given by
 Eq.~(\ref{eq9}), in the form of $\dot{\rho}_\Lambda+3H\rho_\Lambda
 \left(1+w_\Lambda\right)=-Q\not=0$. Yes, we assumed the EoS
 of this fluid $w_\Lambda=-1$ in the second perspective
 considered in Secs.~\ref{sec2}-\ref{sec3}, and
 then $\dot{\rho}_\Lambda=-Q\not=0$. However, due to the
 non-zero $Q$, again one cannot say $\rho_\Lambda=const.$ In
 fact, the two perspectives considered in Sec.~\ref{sec4} and
 Secs.~\ref{sec2}-\ref{sec3} are completely independent. There
 is {\em no} inconsistency in each independent perspective. One
 should not mix up these two different perspectives considered
 in this work, otherwise confusion and misunderstanding might arise.

%============================= acknowledgements ===================================

\section*{ACKNOWLEDGEMENTS}

We thank Prof.~Dominik Schwarz and the anonymous referee for
 quite useful comments and suggestions, which helped us to
 improve this work. We are grateful to Profs.~Rong-Gen~Cai
 and Shuang~Nan~Zhang for helpful discussions. We also thank
 Minzi~Feng, as well as Ya-Nan~Zhou and Shoulong~Li, for
 kind help and discussions. This work was supported in
 part by NSFC under Grants No.~11575022 and No.~11175016.

\renewcommand{\baselinestretch}{1.02}

%============================= references ==================================

\end{document}